\documentclass{article}
\usepackage{graphicx}
\usepackage{amsmath}

\renewcommand{\thetable}{\Roman{table}}

\begin{document}
\baselineskip 18pt

\begin{center}
{\Large{\bf Solar models and solar neutrinos}}
\vskip5mm
J. N. Bahcall$^{1,}$\footnote{Email address: jnb@sns.ias.edu}
\vskip3mm
\mbox{}$^{1}$ Institute for Advanced Study, School of Natural Sciences,
  Princeton, NJ 08540, USA
\bigskip

\begin{abstract}
I provide a summary  of the current theoretical knowledge of solar
neutrino fluxes as derived from precise solar models.

\end{abstract}
\end{center}
\newpage

\section{Introduction}
\label{sec:introduction}

I summarize in this talk the present state of solar model research
as the subject relates to solar neutrino investigations. This is
not a review talk. I will focus on the latest developments. [In a
few places, I will add comments within square brackets on results
that have been obtained since the symposium took place.]

To establish the appropriate context for our discussion, I note
that measurements of the sound speed inside the Sun (via a
technique called helioseismology, which is similar to terrestrial
seismology) agree with solar model predictions to a
root-mean-square accuracy of better than 0.1\%. This excellent
agreement between the measured and predicted sound speeds
established as early as 1996 that the solution of the "solar
neutrino problem" lay in new particle physics, not new
astrophysics~\cite{helioseismology}.

There are two specific  reasons for continuing to study the solar
neutrino fluxes that are predicted by precise solar models. First,
the model predictions must be refined (uncertainties reduced) in
order to use optimally the neutrino observations to provide
information about the interior of the Sun and about some of the
parameters used in solar and stellar models. Second, the
theoretical knowledge of the neutrino fluxes can be used, in
combination with solar neutrino experiments, to refine neutrino
parameters such as $\Delta m^2_{21}$ and $\tan^2 \theta_{12}$
(see, e.g.,~\cite{lisi}).

More generally, we want to use solar models and solar observations
as a laboratory for testing and improving the theory of stellar
evolution, the theory of how stars shine and evolve. The Sun is
the nearest star to Earth and we have much more information about
the Sun than about any other star. We can use the comparison
between the predictions of precise solar models and the
observations of solar properties to explore and refine the theory
of stellar evolution.  Almost every branch of astronomy and
astrophysics uses stellar evolution theory, in one way or another,
to interpret observations of distant stars and the elements they
produce.

Recent measurements of the surface chemical composition of the Sun
have posed a new 'solar problem', a problem that does not have
significant implications for neutrinos but may be of importance
for the theory of stellar evolution and perhaps for other branches
of astronomy. My own work has been focused recently on this aspect
of solar models and I will comment briefly on the subject toward
the end of my talk.

I begin in Section~\ref{sec:models} by summarizing the standard
solar model predictions of solar neutrino fluxes. I also summarize
the uncertainties in the predicted fluxes. For the first two and a
half decades of the solar neutrino problem, the most critical
theoretical issue was to establish that the uncertainties in the
predictions were less than the discrepancies between the solar
model predictions and the solar neutrino measurements. This goal
was achieved somewhere between 1982 and 1966, depending upon the
degree of skepticism toward astronomical calculations and
measurements of the person making the judgment. Actually, for
decades most physicists and a surprisingly large number of
astronomers believed that inaccurate solar model calculations
rather than neutrino properties were responsible for the solar
neutrino problem. This situation changed dramatically with the
publication of the first SNO experimental results in June, 2001,
when it became clear to all that the solution of the solar
neutrino problem was new physics rather than inadequate astronomy.

In Section~\ref{sec:recentdevelopments}, I comment briefly on the
implications of recent measurements of the surface chemical
composition of the Sun. I conclude in Section~\ref{sec:darkmatter}
with a discussion of the role of neutrinos as dark matter, a
subject treated at this symposium by Max Tegmark more extensively
and from a different perspective.

\section{Solar model fluxes}
\label{sec:models}

I base the discussion in this section on the results reported in
the recent paper~\cite{bp04}. Full numerical details of the solar
models, BP04 and BP00 that are discussed below are presented,
together with earlier solar models in this series, at the Web
site: http://www.sns.ias.edu/$\sim$jnb . In particular, the
primordial hydrogen and helium mass fractions for the BP04 model
discussed below  are, respectively, $X_0 = 0.7078$ and $Y_0 =
0.2734$, while the current surface abundances are $X_S = 0.7397$
and $Y_S = 0.2434$.

\subsection{Fluxes from different solar models}
\label{subsec:fluxes}

\begin{table}[!t]
\caption{Predicted solar neutrino fluxes from solar models. The
table presents the predicted fluxes, in units of $10^{10}(pp)$,
$10^{9}({\rm \, ^7Be})$, $10^{8}(pep, {\rm ^{13}N, ^{15}O})$,
$10^{6} ({\rm \, ^8B, ^{17}F})$, and $10^{3}(hep)$ ${\rm
cm^{-2}s^{-1}}$. Columns 2-4 show BP04, BP04+, and the previous
best model BP00~\cite{bp00}. Columns 5-7 present the calculated
fluxes for solar models that differ from  BP00  by an improvement
in one set of input data: nuclear fusion cross sections (column
5), equation of state for the solar interior (column 6), and
surface chemical composition for the Sun (column 7). Column~8 uses
the same input data as for BP04 except for a recent report of the
$^{14}$N + p fusion cross section. References to the improved
input data are given in the text.  The last two rows ignore
neutrino oscillations and present for the chlorine and gallium
solar neutrino experiments the capture rates in SNU (1 SNU equals
$10^{-36}{\rm ~events ~per~target~atom~per~sec}$). Due to
oscillations, the measured rates are smaller: $2.6 \pm 0.2$ and
$69 \pm 4$, respectively. The neutrino absorption cross sections
and their uncertainties are given in Ref.~\cite{nucrosssections}.
 \protect\label{tab:neutrinofluxes}}
\begin{center}
\begin{tabular}{@{}lccccccc}
\noalign{\smallskip}
\hline
\noalign{\smallskip}
Source&\multicolumn{1}{c}{BP04}&{BP04+}&BP00&Nucl&EOS&Comp&$^{14}$N\\
\noalign{\smallskip}
\hline
\noalign{\smallskip}
$pp$&5.94$(1 \pm 0.01)$&5.99 &5.95&5.94&5.95&6.00&5.98\\
$pep$&1.40$(1 \pm 0.02) $&1.42&1.40&1.40&1.40&1.42&1.42\\
$hep$&$7.88 (1 \pm 0.16)$&8.04&9.24&7.88&9.23&9.44&7.93\\
${\rm ^7Be}$&$4.86 (1 \pm 0.12)$&4.65&4.77&4.84&4.79&4.56&4.86\\
${\rm ^8B}$&5.82$(1 \pm 0.23)$&5.28&5.05&5.79&5.08&4.62&5.77\\
${\rm ^{13}N}$&$5.71(1~~^{+0.37}_{-0.35}) $&4.06&5.48&5.69&5.51&3.88&3.23\\
${\rm ^{15}O}$&$5.03(1~~^{+0.43}_{-0.39}) $&3.54 &4.80&5.01&4.82&3.36&2.54\\
${\rm ^{17}F}$&$5.91(1~~^{+0.44}_{-0.44}) $&3.97&5.63&5.88&5.66&3.77&5.85\\
\noalign{\smallskip} \hline \noalign{\smallskip}
Cl&$8.5^{+1.8}_{-1.8}$&7.7&7.6&8.5&7.6&6.9&8.2\\
Ga&131$^{+12}_{-10}$&126&128&130&129&123&127\\
\noalign{\smallskip}
\hline
\end{tabular}
\end{center}
\end{table}

Table~\ref{tab:neutrinofluxes}, taken from Ref.~\cite{bp04}, gives
the calculated solar neutrino fluxes for a series of solar models
calculated with different plausible assumptions about the input
parameters. The range of fluxes shown for these models illustrates
the systematic uncertainties in calculating solar neutrino fluxes.
The second (third) column, labeled BP04 (BP04+), of
Table~\ref{tab:neutrinofluxes} presents the current best solar
model calculations for the neutrino fluxes. The uncertainties in
the calculated neutrino fluxes are given in column~2. The model
BP04 utilizes the older heavy element abundances on the surface of
the Sun as summarized in the paper by Grevesse and
Sauval (1998)~\cite{oldcomp}. The BP04+ model uses recently
determined heavy element abundances~\cite{newcomp}, which are
discussed in more detail in Section~\ref{sec:recentdevelopments}.

\begin{figure}[!t]
\begin{center}
\includegraphics [width=4in,angle=270]{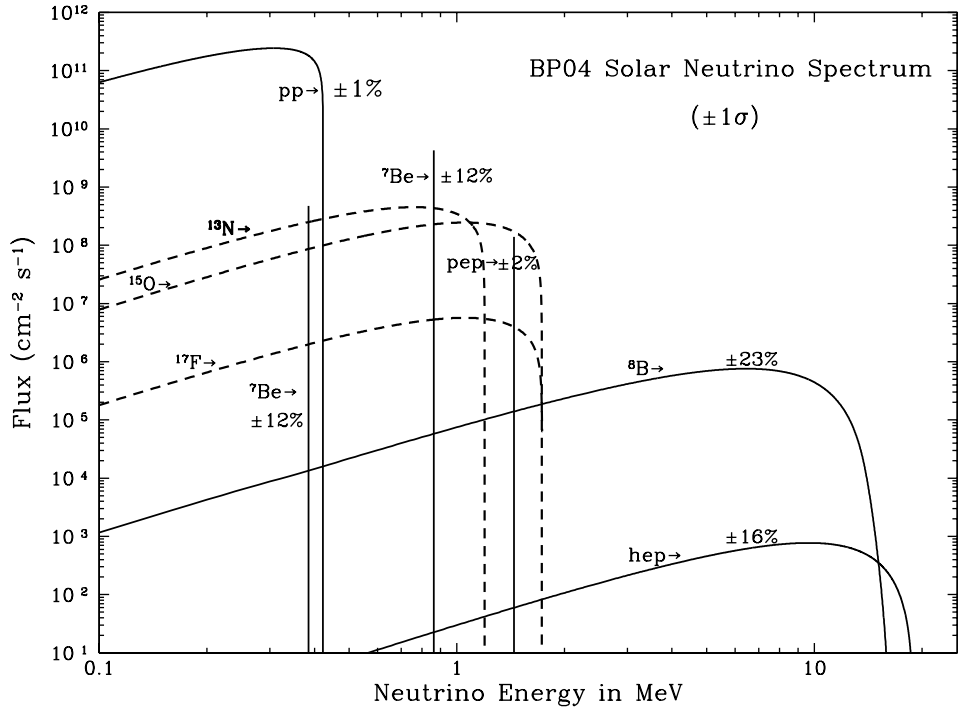}
\end{center}
\caption{The predicted solar neutrino energy spectrum. The figure
shows the energy spectrum of solar neutrinos predicted by the BP04
solar model~\cite{bp04}.  For continuum sources, the neutrino
fluxes are given in number per ${\rm cm^{-2} sec^{-1} MeV^{-1}}$
at the Earth's surface. For line sources, the units are number per
${\rm cm^{-2} sec^{-1}}$. The total theoretical uncertainties for
the neutrinos in the p-p chain are taken from column 2 of
Table~\ref{tab:neutrinofluxes} and are shown for each source. In
order not to complicate the figure, I have omitted the very large
uncertainties for the difficult-to-detect CNO neutrino fluxes (see
Table~\ref{tab:neutrinofluxes}). \label{fig:bp04} }
\end{figure}

Figure~\ref{fig:bp04} presents the neutrino energy spectrum
predicted by the BP04 solar model for the most important solar
neutrino sources.

The model BP04+ was calculated with the use of new input data for
the equation of state, nuclear physics, and solar composition. The
model BP04, the currently preferred model, is the same as BP04+
except that BP04 does not include the most recent analyses of the
solar surface composition~\cite{newcomp}. My collaborators and I
prefer the model BP04 over the model BP04+ because the lower heavy
element abundance used in calculating BP04+ causes the calculated
sound speeds and the depth of the solar convective zone to
conflict with helioseismological measurements~\cite{bp04,helio04}.

The error estimates, which are the same for the three models
labeled BP04, BP04+, and $^{14}$N in
Table~\ref{tab:neutrinofluxes}) include the recent composition
analyses. The differences between the neutrino fluxes calculated
for the solar models BP04 and BP04+ are all well within the
estimated uncertainties in the calculated fluxes. This is one way
of seeing that the 'new solar problem' associated with the lower
heavy element abundances that have been reported recently does not
affect solar neutrino questions in a significant way.

Column four of Table~\ref{tab:neutrinofluxes} presents the fluxes
calculated using the preferred solar model, BP00~\cite{bp00}, that
was posted on the archives in October 2000 (prior to the first
reported SNO measurement). The BP04 best-estimate neutrino fluxes
and their uncertainties have not changed markedly from their BP00
values despite refinements in input parameters. The only exception
is the CNO flux uncertainties which have almost doubled   due to
the larger systematic uncertainty in the surface chemical
composition estimated in this paper.

The columns 5-7 of Table~\ref{tab:neutrinofluxes} describe
improvements in the input data relative to BP00, i.e., relative to
our best standard solar model constructed in 2000. Quantities that
are not discussed here are the same as for BP00. Each class of
improvement is represented by a separate column, columns 5-7. The
magnitude of the changes between the fluxes listed in the
different columns of Table~\ref{tab:neutrinofluxes} are one
measure of the sensitivity of the calculated fluxes to the input
data.

 Column~5 contains the fluxes
computed for a solar model that is identical to BP00 except that
improved values for direct measurements of the
$^7$Be(p,$\gamma$)$^8$B cross section~\cite{b8,pphep}, and the
calculated p-p and hep cross sections~\cite{pphep}. The reactions
that produce the $^8$B and hep neutrinos are rare; changes in
their production cross sections only affect, respectively, the
$^8$B and hep fluxes. The 15\% increase in the calculated $^8$B
neutrino flux, which is primarily due to a more accurate cross
section for $^7$Be(p,$\gamma$)$^8$B, is the only significant
change in the best-estimate fluxes.

The fluxes in Column~6 were calculated using a refined equation of
state, which includes relativistic corrections and a more accurate
treatment of molecules~\cite{eos}. The equation of state
improvements between 1996 and 2001, while significant in some
regions of parameter space, change all the solar neutrino fluxes
by less than 1\%. Solar neutrino calculations are insensitive to
the present level of uncertainties in the equation of state.

The most important changes in the astronomical data since BP00
result from new analyses of the surface chemical composition of
the Sun. The input chemical composition affects the radiative
opacity and hence the physical characteristics of the solar model,
and to a lesser extent the nuclear reaction rates. New values for
C,N,O,Ne, and Ar have been derived~\cite{newcomp} using
three-dimensional rather than one-dimensional atmospheric models,
including hydrodynamical effects, and paying particular attention
to uncertainties in atomic data and observational spectra. The new
abundance estimates, together with the previous best-estimates for
other solar surface abundances~\cite{oldcomp}, imply a ratio of
heavy elements to hydrogen by mass of $Z/X = 0.0176$, much less
than the previous value of $Z/X = 0.0229$~\cite{oldcomp}. Column~7
gives the fluxes calculated for this new composition mixture. The
largest change in the neutrino fluxes for the  p-p chain is the
9\% decrease   in the predicted $^8$B neutrino flux. The N and O
fluxes are decreased by much more, $\sim 35 \%$, because they
reflect directly the inferred C and O abundances. [Subsequent to
the completion of the calculations described here, additional
changes in the recently-determined heavy element abundances have
been made and are reported in Ref.~\cite{AGS05}. These additional
changes are sufficiently small that they do not affect any of the
conclusions expressed in the present review.]

The CNO nuclear reaction rates are less well determined than the
rates for the more important (in the Sun) p-p
reactions~\cite{adelberger}. The rate for
$^{14}$N(p,$\gamma$)$^{15}$O is poorly known, but is important for
calculating CNO neutrino fluxes.  Extrapolating to the low
energies relevant for solar fusion introduces a large uncertainty.
Column 8 gives the neutrino fluxes calculated with input data
identical to BP04  except for the  cross section factor $S_0({\rm
^{14}N + p}) = 1.77 \pm 0.2\, {\rm keV~b}$ that is about half the
current best-estimate;  this value assumes a particular R-matrix
fit to the experimental data~\cite{new14n}. The recent LUNA
measurement~\cite{luna}  of the  cross section factor $S_0({\rm
^{14}N + p})$ provides a firm experimental basis for the assumed
smaller value of the  cross section factor.  The p-p cycle fluxes
are changed by only $\sim 1$\%, but the $^{13}$N and $^{15}$O
neutrino fluxes are reduced by $ 40\%-50$\% relative to the BP04
predictions. CNO nuclear reactions contribute 1.6\% of the solar
luminosity in the BP04 model and only 0.8\% in the model with a
reduced $S_0({\rm ^{14}N + p})$.

\subsection{Flux uncertainties}
\label{subsec:fluxuncertainties}

Table~\ref{tab:uncertainties}, also taken from Ref.~\cite{bp04},
shows the individual contributions to  the flux uncertainties.
These uncertainties are useful in deciding how accurately we need
to determine  a given input parameter in order to improve the
overall accuracy of the calculated neutrino fluxes. Moreover, the
theoretical flux uncertainties continue to play a significant role
in some determinations of neutrino parameters from solar neutrino
experiments (see, e.g., Ref.~\cite{postkamland}).

\begin{table}[!t]
\caption[]{Principal sources of uncertainties in calculating solar
neutrino fluxes.
 Columns 2-5 present the fractional uncertainties in the neutrino fluxes from laboratory
 measurements of, respectively, the $^3$He-$^3$He, $^3$He-$^4$He, p-$^7$Be, and p-$^{14}$N
 nuclear fusion reactions. The last four columns, 6-9, give, respectively, the fractional uncertainties
 due to  the calculated radiative opacity, the calculated rate of element diffusion,
 the measured solar luminosity, and the measured heavy element to hydrogen ratio.\protect\label{tab:uncertainties}}
\begin{center}
\begin{tabular}{lccccccccc}
\noalign{\smallskip}
\hline
\noalign{\smallskip}
Source&\multicolumn{1}{c}{3-3}&3-4&1-7&1-14&Opac&Diff&$L\odot$&Z/X\\
\noalign{\smallskip}
\hline
\noalign{\smallskip}
$pp$&0.002&0.005&0.000&0.002&0.003&0.003&0.003&0.010\\
$pep$&0.003&0.007&0.000&0.002&0.005&0.004&0.003&0.020\\
$hep$&0.024&0.007&0.000&0.001&0.011&0.007&0.000&0.026\\
${\rm ^7Be}$&0.023&0.080&0.000&0.000&0.028&0.018&0.014&0.080\\
${\rm ^8B}$&0.021&0.075&0.038&0.001&0.052&0.040&0.028&0.200\\
${\rm ^{13}N}$&0.001&0.004&0.000&0.118&0.033&0.051&0.021&0.332\\
${\rm ^{15}O}$&0.001&0.004&0.000&0.143&0.041&0.055&0.024&0.375\\
${\rm ^{17}F}$&0.001&0.004&0.000&0.001&0.043&0.057&0.026&0.391\\
\noalign{\smallskip}
\hline
\end{tabular}
\end{center}
\end{table}

Columns~2-5 present the fractional uncertainties from the nuclear
reactions whose measurement errors are most important for
calculating neutrino fluxes. Unless stated otherwise, the
uncertainties in the nuclear fusion cross sections are taken from
Ref.~\cite{adelberger}.

The measured rate of the $^3$He-$^3$He reaction, which
 changed by a factor of 4 after the first solar
model calculation of the solar neutrino flux~\cite{series}, and
the measured rate of the $^7$Be + p reaction, which for most of
this series has been the dominant uncertainty in predicting the
$^8$B neutrino flux, are by now very well determined. If the
current published systematic uncertainties for the $^3$He-$^3$He
and $^7$Be + p reactions are correct,then the uncertainties in
these reactions no longer contribute in a crucial way to the
calculated theoretical uncertainties (see column~2 and column~4 of
Table~\ref{tab:uncertainties}). This felicitous situation is the
result of an enormous effort extending over four decades, and
represents a great collective triumph, for the nuclear physics
community.

 At the present time, the most important nuclear
physics uncertainty in calculating solar neutrino fluxes is  the
rate of the $^3$He-$^4$He reaction (see column~3 of
Table~\ref{tab:uncertainties}).  The systematic uncertainty in the
rate of $^3$He($^4$He, $\gamma$)$^7$Be reaction (see
Ref.~\cite{adelberger})  causes an 8\% uncertainty in the
prediction of both the $^7$Be and the $^8$B solar neutrino fluxes.
It is scandalous that there has not been any progress in the past
15 years in measuring this rate more accurately. [A very recent
precise measurement has been reported for this reaction at
center-of-mass energies above 400 keV~\cite{hass}, but so far
there have not been any other measurements that could test the
accuracy of this result.]

For $^{14}$N(p,$\gamma$)$^{15}$O, we have continued to use in
Table~\ref{tab:uncertainties} the uncertainty given in
Ref.~\cite{adelberger}, although the recent reevaluation in
Ref.~\cite{new14n} suggests that the uncertainty could be somewhat
larger (see column~7 of Table~\ref{tab:neutrinofluxes}).

The dominant uncertainty, 15.1\% ($1\sigma$), for the hep neutrino
flux results from the difficult theoretical calculation of the low
energy cross section factor for this
reaction~\cite{hepuncertainty}.

The uncertainties due to the calculated radiative opacity and
element diffusion, as well as the measured solar luminosity
(columns 6-8 of Table~\ref{tab:uncertainties}), are all moderate,
non-negligible but not dominant. For the $^8$B and CNO neutrino
fluxes, the uncertainties that are due to the radiative opacity,
diffusion coefficient, and solar luminosity are all in the range
2\% to 6\%.

The surface composition of the Sun is the most problematic and
important source of uncertainties. Systematic errors dominate: the
effects of line blending, departures from local thermodynamic
equilibrium, and details of the model of the solar atmosphere. In
the absence of detailed information to the contrary, it is  assumed
that the uncertainty in all important element abundances is
approximately the same. The $3\sigma$ range of $Z/X$ is defined as
the spread over all modern determinations (see
Refs.~\cite{bp00,series,book}), which  implies that at present
$\Delta (Z/X)/(Z/X) = 0.15 ~(1\sigma)$, 2.5 times larger than the
uncertainty adopted in discussing the predictions of the model
BP00~\cite{bp00}. The most recent uncertainty quoted for oxygen,
the most abundant heavy element in the Sun, is similar: 12\%
\cite{newcomp}.

Heavier elements like Fe affect the radiative opacity and hence
the neutrino fluxes more strongly than the relatively light
elements~\cite{bp00}.  This is the reason why the difference
between the fluxes calculated with BP04 and BP04+ (or between BP00
and Comp, see Table~\ref{tab:neutrinofluxes}) is less than would
be expected for the 26\% decrease in $ Z/X$. The abundances that
have changed significantly since BP00 (C, N, O, Ne, Ar) are all
for lighter volatile elements for which meteoritic data are not
available.

 The dominant uncertainty listed in Table~\ref{tab:uncertainties} for
the $^8$B and CNO neutrinos is  the chemical composition,
represented by $Z/X$ (see column~9). The uncertainty ranges from
20\% for the $^8$B neutrino flux to $\sim 35$\% for the CNO
neutrino fluxes. Since the publication of BP00, the best published
estimate for Z/X decreased by $4.3\sigma$ (BP00 uncertainty) and
the estimated uncertainty due to $Z/X$ increased for $^8$B
($^{15}$O) neutrinos by a factor of 2.5 (2.8). Over the past three
decades, the changes have almost always been toward a smaller
$Z/X$. The monotonicity  is surprising since different sources of
improvements have caused successive changes. Nevertheless, since
the changes are monotonic, the  uncertainty estimated from the
historical record is large.

The total $^8$B neutrino flux measured by the neutral current mode
of the SNO experiment~\cite{sno} is
\begin{equation}
\phi(^8{\rm B, SNO}) ~=~ 0.90 \phi(^8{\rm B,
BP04~solar~model})\left[ 1.0 \pm 0.09 \pm 0.23 \right] \, ,
\label{eq:b8snovsstandardmodel}
\end{equation}
where the first uncertainty listed in
equation~(\ref{eq:b8snovsstandardmodel}) is the $1\sigma$
measurement error and the second (larger) uncertainty is the
estimated $1\sigma$ uncertainty in the solar model calculation
(taken from Table~\ref{tab:uncertainties}).  If all the data from
solar neutrino and reactor experiments are combined together (see
figure~\ref{fig:theoryvsexp} below), the above relation
becomes~\cite{postkamland}
\begin{equation}
 \phi(^8{\rm B, SNO}) ~=~ 0.87 \phi(^8{\rm B,
 BP04~solar~model})\left[ 1.0 \pm 0.05 \pm 0.23 \right] \, .
 \label{eq:b8fitvsstandardmodel}
 \end{equation}

The calculated $^8$B neutrino flux~\cite{bp04} agrees with the
measured flux to better than $1\sigma$.  The theoretical
uncertainty is much larger than the uncertainty in the
measurements. The dominant theoretical uncertainty for the $^8$B
neutrino flux is due to uncertainties in the measured surface
heavy element abundances of the Sun.

[For decades, the composition uncertainty has been calculated by
considering the uncertainty in the total heavy element abundance
by mass, Z, or the uncertainty in the total heavy element to
hydrogen ratio, Z/X. In an article now in preparation, Bahcall and
Serenelli 2005, we have made a more refined calculation of the
abundance uncertainties by evaluating separately the abundance
uncertainty of each heavy element. The final result for the total
uncertainty to be used in equation~(\ref{eq:b8snovsstandardmodel})
and equation~(\ref{eq:b8fitvsstandardmodel}) is $\pm 0.16$ instead
of $\pm 0.23$.]

The p-p solar neutrino flux has been measured by combining the
results from all the relevant solar neutrino and reactor
experiments, together with the imposition of the luminosity
constraint~\cite{luminosity}.  The result is~\cite{postneutrino04}
\begin{equation}
\phi({\rm p-p, all~neutrino~experiments}) ~=~ 1.01 \phi({\rm p-p,
BP04~solar~model})\left[ 1.0 \pm 0.02 \pm 0.01 \right] \, ,
\label{eq:ppexperimenttheory}
\end{equation}
where the first uncertainty listed in
equation~(\ref{eq:ppexperimenttheory}) is the $1\sigma$
measurement error and the second (smaller) uncertainty is the
estimated $1\sigma$ uncertainty in the solar model calculation
(taken from Table~\ref{tab:uncertainties}).

\begin{figure}[!t]
\begin{center}
\includegraphics [width=4in,angle=270]{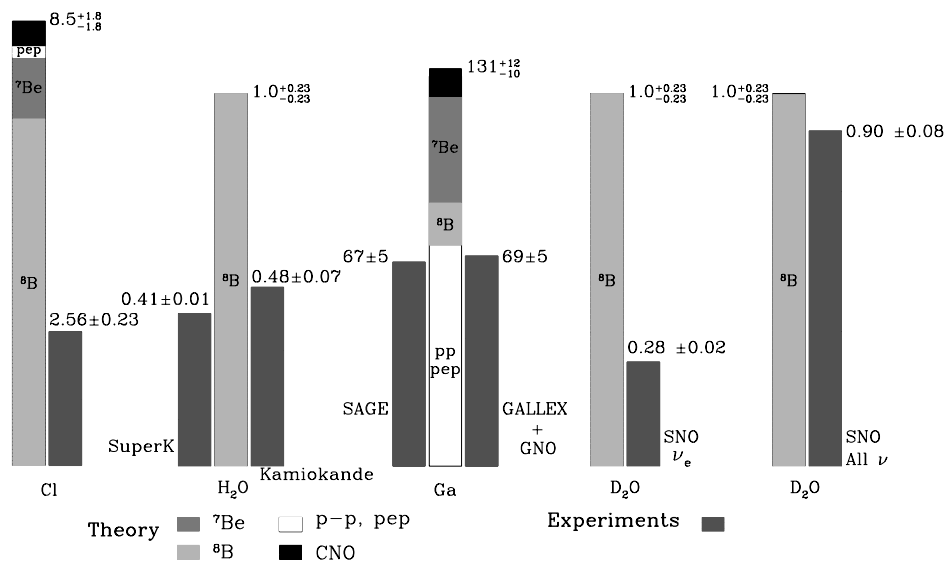}
\end{center}
\caption{Theory versus experiment. The figure compares the
predictions of the standard solar model plus the standard model of
electroweak interactions with the measured rates in all solar
neutrino experiments. The solar neutrino experimental results are
discussed in talks by Y. Suzuki and A. McDonald in this
proceeding. \label{fig:theoryvsexp} }
\end{figure}

\section{Recent developments regarding the solar surface
composition}
\label{sec:recentdevelopments}

In the last several years, determinations of the solar surface
abundances of heavy elements have become more refined and detailed
~\cite{newcomp,AGS05,lodders03}. These recent determinations yield
significantly lower values than were previously adopted (e.g., by
Grevesse and Sauval 1998~\cite{oldcomp}) for the abundances of the
volatile heavy elements: C, N, O, Ne, and Ar. However, these
recent abundance determinations also lead to solar models that
disagree with helioseismological measurements~(\cite{bp04,helio04}).

Standard solar models constructed with the newly-determined heavy
element abundances predict incorrectly the solar sound speeds, the
depth of the solar convective zone, and the surface helium
abundance (see especially Ref.~\cite{helio04}). The discrepancies
occur in the temperature region below the solar convective zone,
$2\times 10^6$K to $4.5 \times 10^6$K~\cite{helio04}. In this
temperature domain, the volatile heavy elements, C, N, O, Ne, and
Ar, are partially ionized and their abundances significantly
affect the radiative opacities. Fortunately, neutrinos are
produced in the deep solar interior not in this outer region,
0.45$R_\odot$ to $0.72R_\odot$, where the discrepancies with
helioseismology exist. This is the reason why the new abundances
do not affect significantly the calculated neutrino fluxes, i.e.,
the reason why the models BP04 and BP04+ in
Table~\ref{tab:neutrinofluxes} yield similar neutrino fluxes.

 What is wrong? I don't know. One
possibility we have considered is that the standard radiative
opacity needs to be increased (by about 11\%) near the base of the
convective zone. However, detailed and refined recalculations of
the radiative opacity by the Opacity Project collaboration
disfavor~\cite{seaton,badnelletal2004} the suggestion
~\cite{helio04,basu04} that the origin of the discrepancy might be
the adopted opacities rather than the adopted heavy element
abundances. Another possibility that astronomers discuss privately
is that something is systematically wrong with the new abundance
determinations. After all, the old standard abundances lead to
solar models that are in excellent agreement everywhere with the
helioseismological measurements. However, no one has identified a
serious problem with the new abundance measurements. But, it would
certainly be healthy for the field if more than one group were to
undertake a thorough reevaluation of the surface heavy element
abundances of the Sun.

As  we continue to measure more precisely the properties of the
Sun and to compare the results with theoretical solar models, we
continue to learn new things. Nature seems to surprise us each
time. The present conflict engendered by the new heavy element
abundance determinations may have significant implications for
other parts of stellar physics. After all, if we can't get the
abundances or the opacity or something else significant correct
for the Sun, about which we know much more than any other star,
how can we have confidence in our quantitative inferences for
other less well studied star?

\section{Neutrinos as dark matter}
\label{sec:darkmatter}

Neutrinos are the first cosmological dark matter to be discovered.
I cannot resist making a few remarks on this subject, because of
the importance of dark matter to both physics and astronomy.

Solar and atmospheric neutrino experiments show that neutrinos
have mass but these oscillations experiments only determine the
differences between masses, not the absolute values. If we make
the plausible but unproven assumption that the lowest neutrino
mass, $m_1$, is much less than the square root of $\Delta m^2_{\rm
solar}$, then we can conclude that the mass of cosmological
neutrino background is dominated by the mass of the heaviest
neutrino. This heaviest neutrino mass is then determined by
$\Delta m_{\rm atmospheric}^2$. With this assumption the
cosmological mass density in neutrinos is
only~\cite{mcdonald,3nuupdate,hitoshicarlos}

\begin{equation}
\Omega_\nu ~=~ (0.0009 \pm 0.0001)\, , ~~\, m_1 << \sqrt{(\Delta
m^2_{\rm solar})}. \label{eq:omegacosmological}
\end{equation}
Although the mass density given in
Equation~(\ref{eq:omegacosmological}) is small, it is of the same order
of magnitude as the observed mass density in stars and gas.

The major uncertainty in determining by neutrino experiments the
value of $\Omega_\nu$ is the unknown value of the lowest neutrino
mass.  It is  possible that  neutrino masses are nearly degenerate
and cluster around the highest mass scale allowed by direct
beta-decay experiments.  If, for example, all neutrino masses are
close to 1 eV, then $\Omega_\nu({\rm 1~ev}) \sim 0.03$, which
would be cosmologically significant, but would not explain the
bulk of the dark matter.

More sensitive neutrino beta-decay experiments and neutrinoless
double beta-decay experiments offer the best opportunities for
determining the mass of the lowest mass neutrino and hence
establishing the value of $\Omega_\nu$ from purely laboratory
measurements.

For most of the period of 1968-2001, between the first suggestion
of a solar neutrino problem and its final resolution as indicating
neutrino oscillations, the great majority of physicists believed
that solar neutrino experiments indicated the need for
improvements in the solar model, not in the standard model of
particle physics. In part, this was a matter of aesthetics; the
standard model of particle physics was beautiful and
experimentally successful.  Why mess up this beautiful theory with
an ugly neutrino mass? By contrast with the elegant laboratory
experiments that confirmed the standard electroweak model, the
interior of the appeared Sun remote and complicated. Also, the
belief in inadequate astrophysics rather than incomplete physics
was, I believe, partially a result of a lack of familiarity with
the accuracy of the solar model calculations. For some of the
physicists that believed in solar neutrino oscillations for
particle physics reasons (see the discussion by Pierre Ramond in
this volume), the most attractive possibility was that neutrino
masses accounted for the missing dark matter. One could argue
persuasively, using Ockham's razor, that this was the only
`natural' value. Unfortunately, as has happened so often in the
history of weak interaction physics, our aesthetic principles were
violated. The neutrino has a mass, but the neutrino mass does not
account for nearly all of the dark matter if we adopt the most
plausible mass hierarchy (see
equation~\ref{eq:omegacosmological}).

\end{document}